\documentstyle[preprint,eqsecnum,aps,epsf]{revtex}
\newif\iftightenlines\tightenlinesfalse
\tightenlines\tightenlinestrue

\def\eslt{\not\!\!{E_T}}
\def\to{\rightarrow}

\def\te{\tilde e}

\def\tb{\tilde b}

\def\tst{\tilde t}
\def\ttau{\tilde \tau}
\def\tmu{\tilde \mu}
\def\tg{\tilde g}
\def\tnu{\tilde\nu}

\def\tw{\widetilde W}

\def\tz{\widetilde Z}
\begin{document}
\draft
\preprint{\vbox{\baselineskip=14pt%
   \rightline{FSU-HEP-030101}
   \rightline{UH-511-983-01}
   \rightline{IFIC/01-13}
}}
\title{Impact of Muon Anomalous Magnetic Moment\\
on Supersymmetric Models}
\author{Howard Baer$^1$, Csaba Bal\'azs$^2$, Javier Ferrandis$^3$
and Xerxes Tata$^2$}
\address{
$^1$Department of Physics,
Florida State University,
Tallahassee, FL 32306 USA
}
\address{
$^2$Department of Physics and Astronomy,
University of Hawaii,
Honolulu, HI 96822, USA
}
\address{
$^3$ Instituto de Fisica Corpuscular -- C.S.I.C.
 -- Universitat de Val\`encia  \\
 Ed. de Institutos de Paterna -- Apartado de Correos 22085 -
 46071 Val\`encia, Spain
}
\date{\today}
\maketitle
\begin{abstract}

The recent measurement of $a_\mu =\frac{g_\mu -2}{2}$ by the E821
Collaboration at Brookhaven deviates from the quoted Standard Model (SM)
central value prediction by $2.6\sigma$. The difference between
SM theory and experiment may be easily accounted for in a variety
of particle physics models employing weak scale supersymmetry (SUSY).
Other supersymmetric models are distinctly disfavored.
We evaluate $a_\mu$ for various supersymmetric models, including
minimal supergravity (mSUGRA), Yukawa unified $SO(10)$ SUSY GUTs,
models with inverted mass hierarchies (IMH),
models with non-universal gaugino masses,
gauge mediated SUSY breaking models (GMSB),
anomaly-mediated SUSY breaking models (AMSB) and
models with gaugino mediated SUSY breaking (inoMSB).
Models with Yukawa coupling unification or multi-TeV
first and second generation scalars are disfavored by the $a_\mu$
measurement.

\end{abstract}

\medskip



\section{Introduction}

%
%

Recently, the Brookhaven E821 experiment has announced a new measurement
of the anomalous magnetic moment of the muon\cite{e821}.  The measured
result deviates by $2.6\sigma$ from the central value of the quoted
Standard Model (SM) prediction\cite{cm}: $a_\mu (exp)-a_\mu (SM)
=43(16)\times 10^{-10}$, where $a_\mu =\frac{g_\mu -2}{2}$ (see also
Ref. \cite{nar}).  The largest error on the SM calculation arises from
the hadronic vacuum polarization loops.  These loops are included via
dispersion integrals involving the rate for $e^+e^-\to hadrons$, with
the largest contribution coming from the region around the $\rho (770)$
resonance.\footnote{We note that other evaluations of the hadronic
vacuum polarization contribution (primarily because of a larger
uncertainty) can lead to a SM $a_\mu$ prediction in accord with
experiment: see Ref. \cite{yndurain}.  The recent evaluation by Davier
and H\"ocker\cite{dh} cited in Ref's \cite{e821,cm} uses the much more
precise tau decay data to reduce the uncertainty in the $e^+e^- \to
\pi^+\pi^-$ cross section. This is a potential source of theoretical
uncertainty since the tau decays purely via the weak isospin 1 channel,
while the photon also has an isospin zero component.}  It is anticipated
that improved measurements of these low energy cross section
measurements will reduce the uncertainty in the SM prediction in the
near future, without recourse to tau decay data. Furthermore, additional
data already taken by the E821 collaboration should soon reduce the
experimental uncertainty in $a_\mu $ by about a factor of two.  If the
difference between SM prediction and experimental measurement of $a_\mu$
is maintained and sharpened, then a clear signal for physics beyond the
SM will be obtained.

The error bars on $a_\mu$ are now small
enough to be sensitive to electroweak loop effects, and other effects
of the same order of magnitude.
Weak scale supersymmetry is an especially well motivated extension of
the SM in which contributions to $a_{\mu}$ from loops with
supersymmetric (SUSY) particles naturally have a magnitude comparable to
electroweak effects. Thus the Brookhaven experiment can potentially probe
weak scale supersymmetry. Supersymmetric contributions to $a_\mu$
have been calculated previously\cite{older,oldergrav,old,moroi}, including a number of
very recent papers addressing the new E821 result
\cite{kane,fm,gondolo,nath,moroi2,hisano,ellis,arnowitt,ew,choi,kim,martin,ko}.
The supersymmetric contributions to $a_\mu$ involve chargino-sneutrino
loops and also neutralino-smuon loops.  For $\tan\beta\equiv {v_u\over
v_d}$ not too small, the supersymmetric contributions yield\cite{moroi}
\begin{equation}
\Delta a_\mu^{SUSY}\propto \frac{m_\mu^2\mu M_i\tan\beta}{M_{SUSY}^4},
\end{equation}
where $M_i$ ($i=1,2$) is a gaugino mass, and $M_{SUSY}$ is a characteristic
sparticle mass scale.
Then $\Delta a_\mu^{SUSY}$ grows with $\tan\beta$, and for models
with a positive gaugino mass, has the
same sign as the superpotential Higgs mass term $\mu$.

In this paper, we present calculations of the
contribution to $a_\mu$ from a variety of supersymmetric models.
Some of these models overlap with those recently examined in
Ref. \cite{fm,nath,moroi2,ellis,choi}, while others are new.
In addition, we compare explicitly where possible with
reach projections of Run 2 of the Fermilab Tevatron $p\bar p$ collider
and the CERN Large Hadron Collider (LHC).

The supersymmetric models considered can be organized according to
the assumed mechanism for communication of supersymmetry breaking
from the hidden sector to the visible sector. First, we consider models
with SUSY breaking communicated via gravitational interactions.
These include the minimal supergravity model (mSUGRA), with
universality of soft SUSY breaking terms at the Grand Unification Theory (GUT)
scale. In addition, we also consider minimal $SO(10)$ models with
a high degree of Yukawa coupling unification\cite{blazek}, models with an
inverted scalar mass hierarchy (proposed to solve the SUSY flavor and CP
problems), and models with non-universal gaugino masses.
Next, we show results for models with gauge-mediated SUSY breaking (GMSB),
for various numbers of messenger fields. We also examine
models with anomaly-mediated SUSY breaking (AMSB) and finally,
models with gaugino-mediated SUSY breaking (inoMSB).

In general, for SUGRA and GMSB models, large CP violating phases
may be present (such phases are suppressed in AMSB and inoMSB models).
Phenomenologically, many of these phases must be small in order to
satisfy constraints such as those coming from measurements of the
neutron and electron electric dipole moments. However, it is possible that
some CP violating phases are large, but still satisfy EDM bounds if
certain amplitude cancellations take place. In this paper, we neglect
the presence of CP violating phases on the $g-2$ calculation;
their effect on SUGRA models is noted in Ref. \cite{in}, and on
general MSSM models in Ref. \cite{kane}.

We end with some broad conclusions in the last section. If the apparent
discrepancy between theory and experiment continues to hold, then
SUSY GUT models with a high degree of Yukawa unification will be
strongly constrained, as will be models which solve the SUSY
flavor and CP problems via a decoupling of the first two generations of
scalars.

\section{Gravity-mediated SUSY breaking models}

In this class of models, SUSY is assumed to be broken in
a hidden sector, consisting of fields which do not interact with usual
particles and their superparners via
SM gauge or Yukawa type interactions. SUSY breaking is communicated to the
visible sector via gravitational interactions.
In general, there is no mechanism to suppress
soft SUSY breaking terms which can lead to flavor changing (FC) and CP
violating processes in conflict with experiment. A common (ad-hoc)
solution is to assume {\it universality} of soft SUSY breaking terms
at the unification scale, which suppresses the unwanted FC
processes.

\subsection{Minimal supergravity model (mSUGRA)}

In this model\cite{sugra}, it is assumed that, at the GUT scale,
all scalars have a common mass $m_0$, all gauginos have a common mass
$m_{1/2}$, and all trilinear soft SUSY breaking terms have a
common value $A_0$. Electroweak symmetry is assumed to be broken
radiatively (REWSB), leading to a model parameter set consisting of
\begin{equation}
m_0,\ m_{1/2},\ A_0,\ \tan\beta,\ sign(\mu ) .
\end{equation}
The superparticle masses and mixings can be calculated via
renormalization group evolution between the scale of grand unification
and the weak scale.
We use the mSUGRA mass calculation embedded in the computer
program ISAJET 7.51 for our results\cite{isajet}. We have adapted
the $\Delta a_\mu^{SUSY}$ calculation of Moroi\cite{moroi} for
use with the ISAJET code.

Our results for the mSUGRA model are shown in Fig. \ref{msugra}.  We
plot only parameter planes with $\mu >0$, since the opposite sign of
$\mu$ almost always gives negative contributions to $\Delta
a_\mu^{SUSY}$.  The solid shaded regions are excluded by {\it i}) a lack
of appropriate REWSB, or {\it ii}) a lightest SUSY particle that is not
the lightest neutralino ($\tz_1$), or {\it iii}) by the experimental
lower limits from LEP2 that $m_{\tw_1}>100$ GeV, $m_{\te_1}>100$ GeV and
$m_{\ttau_1}>76$~GeV.  The region below the thick solid contour has a
lightest Higgs mass $m_h<113.5$ GeV, in apparent discord with recent
results\cite{igo} from LEP2 on searches for SM Higgs bosons. The dots
shaded regions are found to have a value of $\Delta a_\mu^{SUSY}$ within
$2\sigma$ of the E821 result: {\it e.g.} $11\times 10^{-10}< \Delta
a_\mu^{SUSY}< 75\times 10^{-10}$.  The regions below the dashed contours
are accessible to SUSY searches via the isolated trilepton or $\eslt$
signals at Run 2 of the Tevatron with 25 fb$^{-1}$ of integrated
luminosity\cite{trilep,run2sugrep}. Finally, the region below the
dot-dashed contour is accessible to SUSY searches at the CERN LHC $pp$
collider, assuming 10 fb$^{-1}$ of integrated luminosity\cite{suglhc}.

In Fig. \ref{msugra}{\it a}), we show the $m_0\ vs. m_{1/2}$ plane for
$\tan\beta =3$, and for $A_0=-2m_0$. For larger values of $A_0$, we find
the entire plane to be excluded\footnote{Strictly speaking, the bound
$m_h>113.5$~GeV applies to the SM Higgs boson and needs to be corrected
for the SUSY case. However, over much of the parameter space $h$ is
close to the SM Higgs boson, and we will for simplicity use this as the
limit in the rest of this paper. For this reason, as well as to allow
for some uncertainty in the computation of $m_h$, the reader should allow
some latitude in the interpretation of this contour.} by LEP2 Higgs searches.
At this low
value of $\tan\beta$, the region preferred by the E821 result consists
of a small region in the lower left corner.  The preferred region lies
entirely within the Tevatron Run 2 search region, but is unfortunately
already disfavoured by LEP2 Higgs searches.  In frame {\it b}), we show
the same plane, but for $\tan\beta =10$ and $A_0=0$. In this case, the
LEP2 Higgs bound cuts out a significant portion of the region preferred
by $\Delta a_\mu^{SUSY}$, but a substantial region remains at low $m_0$
where $m_h>113.5$ GeV. Much of this region is beyond the reach of
Tevatron Run 2 experiments, but all of the preferred region is well
within the region accessible to LHC searches.  Of particular interest is
that models with large $m_0$ and low $m_{1/2}$, {\it i.e.} in the focus
point region\cite{focus}, are disfavored, and will be excluded if the
disagreement between theory and experiment persists.  Finally, in frame
{\it c}), we show the same parameter space plane as in {\it b}), except
now for $\tan\beta =35$. The region preferred by $\Delta a_\mu^{SUSY}$
has expanded greatly, reflecting the nearly linear growth of $\Delta
a_\mu^{SUSY}$ with $\tan\beta$. Much of the region is beyond the LEP2
Higgs bound, but all of it is within the reach of the CERN LHC.
In this case, the focus point region {\it is} allowed for large $\tan\beta$.

In several papers, comparisons have been made between the calculated values
of $\Delta a_\mu^{SUSY}$ and the $b\to s\gamma $ decay rate, and the
neutralino relic density $\Omega_{\tz_1} h^2$. It is especially
interesting to note that at large $\tan\beta$ and $\mu <0$,
the mSUGRA model is doubly disfavored by both $\Delta a_\mu^{SUSY}$ and
by $b\to s\gamma$\cite{bsg2}. In addition, at large $\tan\beta$, much of
mSUGRA model parameter space at large values of $m_0$ and $m_{1/2}$
is allowed by bounds on the neutralino relic
density\cite{relic,feng_rd,ellis_rd,arn}.
This is due in large part to $\tz_1\tz_1\to A,H\to b\bar b$
annihilation via very broad $A$ and $H$ poles in the $s$-channel.
It is an important observation that the mSUGRA model can accommodate
all these constraints, and simultaneously respect the bound from
LEP2 Higgs searches.

\subsection{Yukawa unified $SO(10)$ model}

Supersymmetric $SO(10)$ grand unified models are especially attractive
in that they can unify gauge couplings, Yukawa couplings and also matter
particles within a single generation.  In light of
evidence for neutrino mass from observation of atmospheric
neutrinos\cite{superk}, there has been heightened interest in this class
of models\cite{soten}.  $SO(10)$ SUSY GUT models naturally accommodate
light neutrinos via the see-saw mechanism.  Using third generation
fermion masses as inputs, Yukawa unification can be achieved if
$\tan\beta \sim 50$ and $\mu<0$.  In this region of parameter space,
however, REWSB is not possible assuming universal scalar masses at the
GUT scale. Incorporation of $D$-term scalar mass contributions, which
are generically present when $SO(10)$ breaks to a lower rank gauge
group, allow REWSB to occur by imposing GUT scale boundary conditions
where $m_{H_u}<m_{H_d}$\cite{so10}.  In this case, the $D$-terms leave a
characteristic imprint on the entire SUSY particle mass spectrum.  The
model parameter space consists of
\begin{equation}
m_{16},\ m_{10},\ M_D^2, m_{1/2},\ A_0,\ \tan\beta,\ {\rm and}\ sign(\mu )  .
\end{equation}
Here, $m_{16}$ is the common soft SUSY breaking mass (renormalized at
the GUT scale) of all matter scalars, while $m_{10}$ is the
corresponding mass of the Higgs scalars.  $M_D$
parametrizes the magnitude of the $SO(10)$ $D$-terms.
Yukawa coupling unification restricts $\tan\beta \sim 50\pm 4$ and
$\mu<0$.
 The sparticle
mass spectrum, relic density, $b\to s \gamma$ decay rate, and collider
signals have been recently calculated in terms of these parameters in
the second paper of Ref.~\cite{so10}.

One might expect these models to be heavily disfavored by the
E821 result, since Yukawa unification occurs for $\mu <0$, while
a positive value of $\Delta a_\mu^{SUSY}$ occurs for $\mu >0$.
We explicitly display these results in Fig.~\ref{so10}, for
{\it a}) $\tan\beta =47$ and {\it b}) $\tan\beta =50$. We generate
models with values of $m_{16}$ and $m_{10}$ over the range $0-3000$ GeV,
$m_{1/2}: 0-1000$ GeV, $-3000<A_0<3000$ GeV and $M_D<1000$ GeV, and
accept models that satisfy REWSB and phenomenological constraints, as
well as having $t-b-\tau$ Yukawa coupling unification at $M_{GUT}$
to 5\%. In both frames, we see that $\Delta a_\mu^{SUSY}$ is always
less than zero. However, for $\tan\beta =47$, a substantial fraction
of models have $|\Delta a_\mu^{SUSY}|< 5$, so that they lie within $3\sigma$
of the E821 result. These models generally require $m_{16}>1000$ GeV,
so are effectively entering the decoupling regime. In frame {\it b}),
for $\tan\beta =50$, almost all models are excluded unless
$m_{16}>2200$ GeV.
For the higher $\tan\beta$ value, larger values of $M_2$ and
smaller values of $|\mu |$ are generated for Yukawa unified solutions;
in some cases, neutralino loops dominate the amplitude for
$\Delta a_\mu^{SUSY}$.
Overall, for $\tan\beta =50$, there is decoupling at larger values of
$m_{16}$ than in the
$\tan\beta =47$ case.
In these models, generally a reasonable value for $\Omega_{\tz_1}h^2$
can be obtained\cite{so10}. Also,
in these decoupling regimes, the rate for anomalous supersymmetric
contributions to $b\to s\gamma$ decay rate also diminishes. However,
such large values of $m_{16}$ may suffer from problems with
naturalness.


\subsection{Inverted scalar mass hierarchy models}

A possible solution to the SUSY flavor and CP problems arises
by {\it decoupling} the matter scalars in the theory. This involves
setting matter scalar masses towards the 10-100 TeV range, thereby suppressing
all loop induced FCNCs and CP violating processes\cite{masiero}. Such models
seemingly violate constraints from naturalness, but two ways out have
been suggested. In the focus point region of mSUGRA models\cite{focus},
large $m_0$, small $m_{1/2}$ and moderate to large $\tan\beta$
yields a small value for $|\mu |$ and
possibly a small fine tuning of model parameters to achieve REWSB.
An alternative is in models with an inverted scalar mass hierarchy (IMH),
wherein first and second generations have multi-TeV masses, while third
generation scalars have sub-TeV masses: this latter condition
results in models satisfying ``naturalness'', since third generation
scalar masses enter the fine-tuning calculation, while first and
second generation masses are most severely
constrained by flavor and CP violating processes.

\subsubsection{Radiatively driven IMH model}

An intriguing scenario has been proposed in Ref. \cite{rimh}, wherein
the IMH is generated {\it radiatively} by starting with multi-TeV
masses for {\it all} scalars at the GUT scale. At large $\tan\beta$,
and for special soft SUSY breaking boundary conditions, third generation
scalar masses are driven to weak scale values via renormalization
group evolution, while first and second generation scalars remain at multi-TeV
values, owing to their small Yukawa couplings. The proposed model incorporates
$t-b-\tau$ Yukawa coupling unification, and necessarily includes right
neutrino superfields (as in $SO(10)$), and requires the special
($SO(10)$ symmetric) boundary
conditions $A_0^2=2m_{10}^2=4 m_{16}^2$ for the renormalization group
evolution of the soft SUSY breaking parameters.
For consistency of Yukawa coupling unification with REWSB,
$SO(10)$ $D$-terms are again necessary \cite{imh2}.
Thus, the parameter space of the RIMH model consists of
\begin{equation}
m_{16},\ m_{1/2},\ \tan\beta,\ M_D^2,\ M_N,\ {\rm and}\ sign(\mu ).
\end{equation}
The value of $\tan\beta$ is generally large to achieve Yukawa unification,
and the parameter $M_N$ is the scale of the superpotential
singlet neutrino mass term: $10^3\ {\rm TeV}<M_N<M_{GUT}$,
with $M_N\sim 10^{15}$ GeV favored by data from atmospheric neutrinos in
the simplest seesaw model for neutrino masses. Only a modest IMH is possible.
The mass spectra associated with these models, along with expectations for
relic density, $b\to s\gamma $ decay rates and collider signals, have
been discussed in Ref. \cite{imh2,imh3}.

Motivated by the considerations in Ref. \cite{imh3}, we show in
Fig. \ref{imhn}{\it a}) the masses of several sparticles,
and in {\it b}) the associated value of $\Delta a_\mu^{SUSY}$, for
$\mu <0$, $\tan\beta =50$, $M_D=0.2 m_{16}$, $m_{1/2}=0.25 m_{16}$ and
$M_N=1\times 10^7$ GeV. These parameters give a reasonable
IMH while maintaining a calculable sparticle mass spectrum.
In {\it a}), the IMH is displayed by the mass gap between $m_{\tmu_1}$
and $m_{\tst_1}$ and $m_{\tb_1}$. In {\it b}), we see that
$\Delta a_\mu^{SUSY}$ is always negative for this case. However,
for values of $m_{16}>3000$ GeV, the value of
$\Delta a_\mu^{SUSY}$ is within $3\sigma$ of the E821 result, as the model
moves into the decoupling regime.
Such large values of $m_{16}$ appear to be beyond the reach of even the
LHC, at least for an integrated luminosity of 10~$fb^{-1}$ \cite{imh3}.

We have also examined the situation regarding $a_{\mu}$ for $\mu > 0$.
In this case, as discussed in Ref.\cite{imh3} Yukawa couplings do not
unify well, and the unification parameter $R$ typically ranges between 1.6 and
1.9. Here, we have fixed $M_D = 0.2m_{16}$ which tends to give the largest
mass hierarchy \cite{imh3} for the chosen $\tan\beta=50$. The results of our
analysis are shown in Fig.~\ref{imhpos} for {\it a})~$M_N=10^7$~GeV, and
{\it b})~$M_N=10^{15}$~GeV. The shaded regions are excluded because
there is no REWSB, or one of the scalar mass parameters is tachyonic, or
the lightest supersymmetric particle (LSP) is not a neutralino.
To the right of the solid line, the crunch
factor $S$ defined in Ref.\cite{imh3} exceeds 4. The dotted area shows
the $2\sigma$ region preferred by the result of E821 experiment.

In Fig. \ref{imhp}, we show $\Delta a_{\mu}^{SUSY}$ versus $m_{16}$ for
the slice of Fig.~\ref{imhpos}{\it a} with
$m_{1/2}=0.22 m_{16}$ for which the hierarchy tends to be large.
We see that $\Delta a_\mu^{SUSY}$ is within $2\sigma$ of the E821
central value when $m_{16}<1700$ GeV. SUSY signals should readily be
observable throughout this range, which unfortunately is disfavoured by
considerations of relic density and the branching ratio for the $b\to
s\gamma$ decay\cite{imh3}.

\subsubsection{GUT scale IMH model}

In this class of models, it is assumed the IMH already exists
at the GUT scale. The model parameter space is
\begin{equation}
m_0(1),\ m_0(3),\ m_{1/2},\ A_0,\ \tan\beta\ {\rm and}\ sign(\mu ),
\end{equation}
where $m_0(1)$ is the common mass of first and second generation scalars
at $M_{GUT}$, while $m_0(3)$ is the mass of all third generation and
Higgs scalars at $M_{GUT}$. In this case, two-loop contributions
from first and second generation scalars to RG evolution of
third generation scalars helps to drive the latter to
small masses; to avoid tachyons, $m_{1/2}$ must be chosen large
enough. These models have been recently explored in Ref. \cite{mur,gsimh}.
In general, GUT scale IMH models can support much larger first and second
generation scalar masses than RIMH models. This helps to further
suppress FC and CP violating processes, but will also suppress
anomalous contributions to $a_\mu$. For instance, for the two case studies
presented in Table 1 of Ref.~\cite{gsimh}, we find
$\Delta a_\mu^{SUSY}= 1.6\times 10^{-13}$ (case 1) and
$1.8\times 10^{-11}$ (case 2). Hence, these models should give $a_\mu$
values very close to the SM prediction.

\subsection{$SU(5)$ models with non-universal gaugino masses}

In supergravity GUT models based on $SU(5)$ gauge symmetry,
gaugino masses arise from the Lagrangian term
\begin{equation}
{\cal L}\supset \frac{\langle F_\Phi\rangle_{ab}}{M_{Pl}}
\lambda_a\lambda_b
\end{equation}
where the $\lambda_a$ are the gaugino fields, and $\langle F_\Phi\rangle$
is the vacuum expectation value of the $F$ term of a chiral superfield which
transforms as the symmetric product
of two adjoints under $SU(5)$:
\begin{equation}
({\bf 24}{\bf \times}
 {\bf 24})_{\rm symmetric}={\bf 1}\oplus {\bf 24} \oplus {\bf 75}
 \oplus {\bf 200} .
\end{equation}
Universal gaugino masses are obtained only if the superfield $\Phi$ is a
singlet of the GUT group. Higher
dimensional $\Phi$ representations yield non-universal gaugino
masses at the GUT scale: for the ${\bf 24}$, $M_1:M_2:M_3 =-1:-3:2$,
while for the ${\bf 75}$, $M_1:M_2:M_3 =-5:3:1$ and for the ${\bf 200}$,
$M_1:M_2:M_3 =10:2:1$.
Thus, the model parameter space\cite{anderson}
is given by\footnote{Here, we do not consider
the possibility that an arbitrary linear combination of these
irreducible representations is also possible.}
\begin{equation}
m_0,\ M_3^0,\ A_0,\ \tan\beta\ {\rm and}\ sign(\mu ),
\end{equation}
where the $SU(2)_L$ and $U(1)_Y$ gaugino masses can be calculated
in terms of $M_3^0$, the GUT scale $SU(3)_C$ gaugino mass.

These $SU(5)$ models with non-universal gaugino masses will yield weak
scale gaugino mass values very different from the mSUGRA prediction, and
potentially also very different values of the muon anomalous magnetic
moment. We have studied the SUSY contributions to $a_{\mu}$ within this
framework for positive values of $M_2\mu$. The 2$\sigma$ region favored
by the E821 experiment is shown in the $m_0\ vs.\ M_3^0$ plane in
Fig.~\ref{gaugino1} for $\tan\beta=10$ and Fig.~\ref{gaugino2} for
$\tan\beta=35$. We fix $A_0=0$ in both figures.  The frames {\it a}),
{\it b}) and {\it c}) respectively show the cases where the superfield
$\Phi$ transforms as a {\bf 24}, {\bf 75} and {\bf 200} dimensional
representation of $SU(5)$. The corresponding cases for the singlet
$\Phi$ are the mSUGRA cases in Fig.~\ref{msugra}{\it b} and
Fig.~\ref{msugra}{\it c}. Again the shaded region in frames {\it
a}) and {\it c}) is excluded by the same theoretical and experimental
constraints discussed for the mSUGRA case. For the {\bf 75} case in
frame {\it b}) the theory constraints together with $m_{\tw_1}> 100$~GeV
exclude the entire plane. However, in the {\bf 75} and {\bf 200} cases,
the mass gap between the chargino and the LSP is very small, and the LEP
constraint on the chargino mass may have to be reassessed. In these cases, the lighter
neutralinos contain significant higgsino components, and LEP experiments
may also be able to probe $\tz_1\tz_2$ production as discussed in
Ref. \cite{anderson}. In view of this, in frames {\it b}), we have chosen
to include just the theory constraints in the dark shaded region. In the
light shaded region (which covers the rest of the plane in
Fig.~\ref{gaugino2}), $m_{\tw_1} < 85$~GeV. The various lines labelled by
the sparticle type denote contours where $m_{\te_1}=250$~GeV,
$m_{\tw_1}=250$~GeV and $m_{\tg}=1,2$~TeV. In frames {\it c}) the
selectron is heavier than 250~GeV throughout the allowed region, while
the gluino is always lighter than 1~TeV in the allowed range of frame
{\it b}). Finally, the solid lines are contours where $m_{h}=113.5$~GeV,
except in frames {\it b}) where $h$ is lighter than this throughout: in
this case, the solid line labels $m_h=110$~GeV.

The area shaded by dots is the region favoured by the E821 experiment
at the $2\sigma$ level. We see that the {\bf 24} cases are qualitatively
similar to the corresponding mSUGRA cases in Fig.~\ref{msugra}, although
the allowed region does not extend as far up in the GUT scale gluino
mass. The bulk of this region would be accessible at a
500~GeV linear collider for $\tan\beta \sim 10$, but even for the larger
value of $\tan\beta$, a considerable portion of this area will be probed
there. While simulations of LHC signals have not been performed within
this framework, presumably this entire region will be probed by LHC
experiments via the usual multi-jet plus multi-lepton signals. Multi-jet
events with identified $Z$ bosons and $\eslt$ are a characteristic
feature of
this framework\cite{anderson}.

Moving to the {\bf 75} case in frames {\it b}), we see that the bulk
(possibly all) of the parameter plane is already excluded by various
constraints already described. Indeed if the region with charginos up to
85~GeV can be excluded either via chargino searches, or via neutralino
searches, and $h$ can definitely be determined to be larger than 110~GeV
even after incorporating effects of mixing amongst the CP even scalars,
this case will be excluded. In any case, it should be possible to probe
this entire region at future colliders, the near degeneracy of the $\tw_1$
and $\tz_1$ notwithstanding.

Finally, in frames {\it c}), we see that the E821 favoured region
extends over a large fraction of the plane for both values of
$\tan\beta$. In the region to the right of the contour labelled $\tw_1$,
$m_{\tw_1}< 250$~GeV, and reduces to below the experimental bound as we
hit the excluded region. Chargino pair production will thus be
accessible only over parts of the parameter space favoured by E821. We
have checked that the lighter chargino is higgsino-like, or for
relatively small values of $M_3^0$ mixed, throughout the plane. Its
detection may thus be complicated by the fact that the $\tw_1 - \tz_1$
mass gap is small over much of this range. Again, although explicit
LHC simulations have not been performed for this scenario, experiments
at the LHC should be sensitive to the SUSY signal over a considerable
fraction (if not all) of this parameter range.

To get some flavour of the size of the various contributions to $a_{\mu}$,
we present in Table 1 a single case
study for each of the models labeled by the dimensionality of the
$\Phi$ representation. The model parameters are also listed in the table.
The corresponding sparticle spectrum is listed in Table 4 of
Ref.~\cite{jhep}. We recognize that this point may well be excluded by
experimental constraints in some of the cases. Our purpose here is only
to illustrate how the individual contributions depend on the
dimensionality of $\Phi$.
In the table, we show the magnitude of the various SUSY loop
contributions to $\Delta a_\mu^{SUSY}$, along with the total.
In the case of ${\bf 1}$ (universality), we see that the chargino loops
dominate, while the $\tz_2\tmu_i$ ($i=1,2)$
loops are also large, but nearly cancel with each other.
For this model, the $\tz_2$ is mainly wino-like. In the ${\bf 24}$ model,
at the weak scale $M_1$ decreases by a factor of $\sim 2$ relative to the
mSUGRA case, while $M_2$ increases by about
$\sim 1.5$. The loop
contributions involving $\tw_i$ decrease relative the mSUGRA case, but
still dominate the total contribution. In the ${\bf 75}$ model,
$M_2$ increases by about a factor of 3 relative to mSUGRA, while
$M_1$ increases by a factor of about 5. The $\tz_3$ loops form the
dominant neutralino contribution, and the total value of $\Delta a_\mu^{SUSY}$
is further suppressed. Finally, in the ${\bf 200}$ case, $M_2$ increases
by a factor of about 2 at the weak scale relative to mSUGRA,
while $M_1$ increases by about 10. In this case, the $\tz_4$ is largely
wino-like, and it gives the dominant contribution to
$\Delta a_\mu^{SUSY}$.

\section{Gauge-mediated SUSY breaking model}

In recent years, there has been much interest in
supersymmetric models where SUSY breaking is communicated via
gauge interactions\cite{gmsb}.
In these models, SUSY breaking occurs in a hidden sector, but
SUSY
breaking is communicated from the hidden sector to the observable sector
via Standard Model gauge interactions of messenger particles,
which are assumed to occur in $n_5$ complete vector
representations of $SU(5)$ with quantum numbers of $SU(2)$ doublets of
quarks and leptons. The messenger sector mass scale is characterized by $M$.
The soft SUSY breaking masses for the SUSY partners of SM particles are
thus proportional to the strength of their gauge interactions, so that
squarks are heavier than sleptons, while the gaugino masses
satisfy the usual ``grand unification'' mass relations,
though for very different reasons. Within the minimal version of this
framework, the couplings and masses of the sparticles in the observable
sector are determined (at the messenger scale $M$) by the parameter set,
\begin{equation}
\Lambda,M,n_5,\tan\beta,sign (\mu ),C_{grav}.
\label{parset}
\end{equation}
The parameter $\Lambda$ sets the scale of sparticle masses and is the
most important of these parameters. The model predictions for soft-SUSY
breaking parameters at the scale $M$ are evolved to the weak scale. The
parameter $C_{grav} \geq 1$ and enters only into the
partial width for sparticle decays to the gravitino.

In Fig.~\ref{gmsb}, we show the $2\sigma$ region favored by the
E821 data for the minimal GMSB model. Our plots are in the
$\Lambda\ vs.\ \tan\beta$ plane, for $M=3\Lambda$, $\mu >0$
and {\it a}) $n_5=1$, {\it b}) $n_5=2$, {\it c}) $n_5=3$
and {\it d}) $n_5=2$ but with $\mu =0.75 M_1$. In all cases but {\it d}),
the magnitude of $\mu$ is fixed by the REWSB constraint. The solid region
is excluded either by the REWSB constraint, or by $m_{\tz_1}<95$ GeV,
$m_{\te_1}<100$ GeV, $m_{\ttau_1}<76$ GeV
or $m_{\tw_1}<100$ GeV, as indicated by LEP2 searches.
The solid contour denotes where
$m_h=113.5$ GeV. Other mass contours listed are $m_{\tw_1}=250$ GeV,
$m_{\te_1}=250$ GeV, $m_{\ttau_1}=250$ GeV, and $m_{\tg}=1$ and 2
TeV (dot-dashed contours).
The $\tw_1$, $\te_1$ and $\ttau_1$ contours correspond to the
approximate reach for SUSY particles of a Next Linear Collider
operating at $\sqrt{s}=500$ GeV. In frame {\it a}), we see that
very little parameter space is favored for low $\tan\beta$, but considerable
parameter space is favored for large $\tan\beta$. The stars correspond
to the reach in $\Lambda$ of the CERN LHC $pp$ collider for specific GMSB
model lines (assuming 10 fb$^{-1}$ of integrated luminosity),
with the $\tan\beta $ values sampled in Ref. \cite{gmsblhc}.
The large dots denote the corresponding reach for the Tevatron with
25~$fb^{-1}$ of integrated luminosity \cite{yilitev}.
In {\it a}),
at least for low $\tan\beta$, even the Tevatron reach extends beyond the
maximum value of $\Lambda $ favored by the E821 result, while the reach
of the LHC is far beyond.
Similar plots are shown in {\it b}) and {\it c}) but for
larger numbers of messenger fields. Frame {\it d}) is shown for the
special case of a higgsino-like NLSP. For the model lines that have been
analyzed \cite{yilitev}, we see that for the range of parameters
favoured by the E821 data, SUSY signals might well be observable even at
the luminosity upgrade of the Tevatron.

\section{Anomaly-mediated SUSY breaking model}

It has recently been recognized\cite{RanSun}
that there exist loop contributions to sparticle masses
originating in the super-Weyl anomaly, which is always present
in supergravity models when SUSY
is broken. In models without SM gauge singlet superfields that can
acquire a Planck scale vev, the usual supergravity contribution to
gaugino masses is suppressed by an additional factor $\frac{M_{SUSY}}
{M_P}$ relative to $m_{\frac{3}{2}} = M_{SUSY}^2/M_P$, or in higher
dimensional models where the coupling between the observable and hidden
sectors is strongly suppressed, these
anomaly-mediated contribution can dominate.
The gaugino masses turn out to be non-universal, and are
found to be proportional to the respective gauge group $\beta$-functions.
Likewise, scalar masses and trilinear terms are given in terms of
gauge group and Yukawa interaction beta functions.
Slepton squared masses turn out to be negative (tachyonic). A common
fix is to assume an additional contribution $m_0^2$ for all scalars.
The parameter space of the model then consists of
\begin{equation}
m_0,\ m_{3/2},\ \tan\beta\ {\rm and}\ sign(\mu ) .
\end{equation}
In the minimal AMSB model (mAMSB),
the $\tw_1$ and $\tz_1$ are both wino-like, and
nearly mass degenerate, leading to a unique phenomenology\cite{amsbpheno}.

In Fig. \ref{amsb}, we show plots of the mAMSB parameter space via the
$m_0\ vs.\ m_{3/2}$ plane, for $\mu<0$, and for {\it a}) $\tan\beta =3$,
{\it b}) $\tan\beta=10$ and {\it c}) $\tan\beta =35$. For AMSB models,
$\mu <0$ yields $\Delta a_\mu^{SUSY}>0$, since we take the gaugino
masses $M_1$ and $M_2$ to be negative in ISAJET.\footnote{We have
corrected an error in the sign of the $A$-parameters that was present
in ISAJET v7.51. This does not affect the results in
Ref.\cite{amsblhc}.}
Incidently, this sign of
$\mu$ appears to be disfavoured by constraints on $BR(b\to s\gamma)$ as
obtained in Ref.\cite{bsg}.
The solid regions are
excluded by lack of the correct pattern of REWSB, or when $\ttau_1$ is
the LSP, or by $m_{\tw_1}<86$ GeV (from LEP2 chargino searches in the
mAMSB model\cite{lep2amsb}). In frame {\it a}), only a tiny region at
the lower left is within the $2\sigma$ range of $\Delta
a_\mu$ as measured by E821.
Moreover, in this entire plane $m_h<113.5$ GeV. The solid contour denotes where
$m_h=105$ GeV: thus, even allowing for effects of mixing in the Higgs
sector, much of this plane is presumably excluded. The dashed-dotted
contour denotes the reach of the CERN LHC for mAMSB models with 10
fb$^{-1}$ of integrated luminosity\cite{amsblhc}. In frame {\it b}), the
solid contour denotes where $m_h=113.5$ GeV, and the dotted shading denotes
the $2\sigma$ favored region of $\Delta a_\mu^{SUSY}$. Only a small
region has $\Delta a_\mu^{SUSY}>11\times 10^{-10}$, and $m_h>113.5$
GeV. We do not show any dashed-dotted contour as there was no
computation of the LHC reach for this value of $\tan\beta$.  In {\it
c}), for $\tan\beta=35$, we find that a considerable region of parameter
space is favored by the muon anomalous moment, while simultaneously
having $m_h>113.5$ GeV (solid contour). The reach of CERN LHC
encompasses almost the entire favored region.

\section{Gaugino-mediated SUSY breaking model}

A fourth class of models has been recently proposed, based
on extra dimensions with branes, which provides a novel
solution to the SUSY flavor and $CP$ problems\cite{inomsb}.
In this framework,
chiral supermultiplets of the observable sector reside on one brane
whereas the SUSY breaking sector is confined to a different
brane. Gravity and gauge superfields propagate in the
bulk, and hence, directly couple to fields on both the branes. As a
result of their direct coupling to the SUSY breaking sector, gauginos
acquire a mass. The scalar components of the chiral supermultiplets,
however, can acquire a SUSY breaking mass only via their interactions
with gauginos (or gravity) which feel the effects of SUSY breaking: as a
result, these masses are suppressed relative to gaugino masses, and may
be neglected in the first approximation. The same is true for
the $A$ parameters.

To gain a phenomenologically acceptable sparticle mass spectrum,
it is necessary to include additional renormalization group running between
the assumed compactification scale $M_c$ and $M_{GUT}$. At energies
above $M_{GUT}$, it is assumed that either $SO(10)$ or $SU(5)$ grand
unification is valid. Thus, all scalars have masses $m_0=0$ at $M_c$,
but non-zero values at $M_{GUT}$.
The parameter space of the model consists of\cite{jhep}
\begin{equation}
m_{1/2}, M_c, \tan\beta , sign(\mu ).
\label{mgm2}
\end{equation}
If one insists on a high degree of
Yukawa coupling unification, then
one is forced to require $\mu <0$. In this case, $\Delta a_\mu^{SUSY}$
will always be negative. Yukawa coupling unification also highly
restricts the allowed values of $\tan\beta$. Other parameters may be
necessary in addition to the above set depending on the specific model
of grand unification assumed.

In Fig. \ref{inomed},  we assume the inoMSB boundary conditions to be
valid at $M_c=1\times 10^{18}$ GeV, and that minimal $SU(5)$ grand
unification is valid between $M_c$ and
$M_{GUT}\simeq 2\times 10^{16}$ GeV. We use the $SU(5)$
RGEs given in Ref. \cite{jhep}, with Yukawa couplings
$f_t=0.519$, and $f_b=f_\tau =0.277$ at $M_{GUT}$, which are
characteristic of $\tan\beta =35$. Two additional $SU(5)$
Yukawa couplings $\lambda =1$ and $\lambda'=0.1$ are assumed.
Several representative sparticle masses are shown in frame {\it a}),
along with the value of $|\mu|$.
Note that the lower bound on parameter space of $m_{1/2}\simeq 280$ GeV
occurs where $m_{\ttau_1}=m_{\tz_1}$. In frame {\it b}), we show the
corresponding value of $\Delta a_\mu^{SUSY}$. As expected, it is always
negative. However, for $m_{1/2}> 1$ TeV, it lies within the $3\sigma$
bound from E821. In this case, the model is moving towards the region of
unnaturalness.

\section{Summary and conclusions}

The recently reported measurement of $a_\mu$ by the Muon ($g-2$)
Experiment E821 has been interpreted as a harbinger of physics beyond
the Standard Model. This will be so if in fact the quoted SM estimate of
$a_\mu$ and its associated error are verified, and if the experimental
measurement is maintained with a reduced error after the analysis of
the year 2000 data.  If the discrepancy between the measured and theoretical
values of $a_\mu$ persists, then the result will act as a strong
constraint on many forms of new physics, including weak scale
supersymmetric matter.

We presented here the values of $\Delta a_\mu^{SUSY}$ expected in a
variety of supersymmetric models. The quoted discrepancy\cite{e821}
between theory and experiment can easily be accommodated in a variety of
supersymmetric models.
These include SUGRA and GMSB models with $\mu >0$ and
moderate to large $\tan\beta$.

The SUGRA, GMSB and AMSB models at low $\tan\beta$ are disfavored.
In addition, models with a negative value of $\mu M_2$
are disfavored; this includes models that
incorporate a high degree
of Yukawa coupling unification such as simple $SU(5)$ or $SO(10)$
SUSY GUT models, and inoMSB models. Finally, models that
invoke a decoupling solution to the SUSY flavor and CP problems
are disfavored. These include models with an inverted scalar mass
hierarchy, and focus point models at intermediate ({\it but not high})
$\tan\beta$.
Even so, like the SM,
these decoupling models are generally allowed at the $3\sigma$ level.

%
\acknowledgments
We thank F. Paige and J. Feng for discussions and comments.
This research was supported in part by the U.~S. Department of Energy
under contract numbers DE-FG02-97ER41022 and DE-FG03-94ER40833.
J.~Ferrandis was supported by a spanish MEC-FPI grant and by
the European Commission TMR contract HPRN-CT-2000-00148.
%
%
%

\newpage
%
%

\iftightenlines\else\newpage\fi
\iftightenlines\global\firstfigfalse\fi
\def\dofig#1#2{\epsfxsize=#1\centerline{\epsfbox{#2}}}

\begin{table}
\begin{center}
\caption{Loop contributions to $\Delta a_\mu^{SUSY}$ in $SU(5)$ models with
non-universal gaugino masses. Each contribution must be
multiplied by $10^{-10}$. We adopt the parameter space point
$(m_0,\ M_3^0,\ A_0 )=(100,\ 150,\ 0 )$ GeV, with $\tan\beta =5$.
For each case, $\mu >0$ except model ${\bf 24}$, for which $\mu <0$.}
\bigskip
\begin{tabular}{lcccc}
\hline
loop & {\bf 1} & {\bf 24} & {\bf 75} & {\bf 200} \\
\hline
$\tw_1\tnu_\mu$ & 44.5  & 38.1  & 11.6  & 19.4 \\
$\tw_2\tnu_\mu$ & -11.8 & -15.6 & -3.62 & -10.5 \\
$\tz_1\tmu_1$ & 4.58  & -1.57 & 0.55  & -2.84 \\
$\tz_2\tmu_1$ & 21.2  & 8.01  & 0.71  & 0.06 \\
$\tz_3\tmu_1$ & -1.15 & -2.08 & -2.47 & 5.49 \\
$\tz_4\tmu_1$ & -0.25 & 0.07  & -0.04 & 48.5 \\
$\tz_1\tmu_2$ & -7.56 & -1.32 & -1.31 & 0.05 \\
$\tz_2\tmu_2$ & -21.8 & -9.74 & -0.69 & -0.37 \\
$\tz_3\tmu_2$ & -1.49 & -0.77 & 1.84  & -1.75 \\
$\tz_4\tmu_2$ & 6.14  & 5.37  & 0.96  & -33.3 \\
$total$ & 32.4 & 20.4 & 7.53 & 24.7 \\
\label{tab:nonuni}
\end{tabular}
\end{center}
\end{table}
\newpage
%

%
\begin{figure}
\dofig{6.5in}{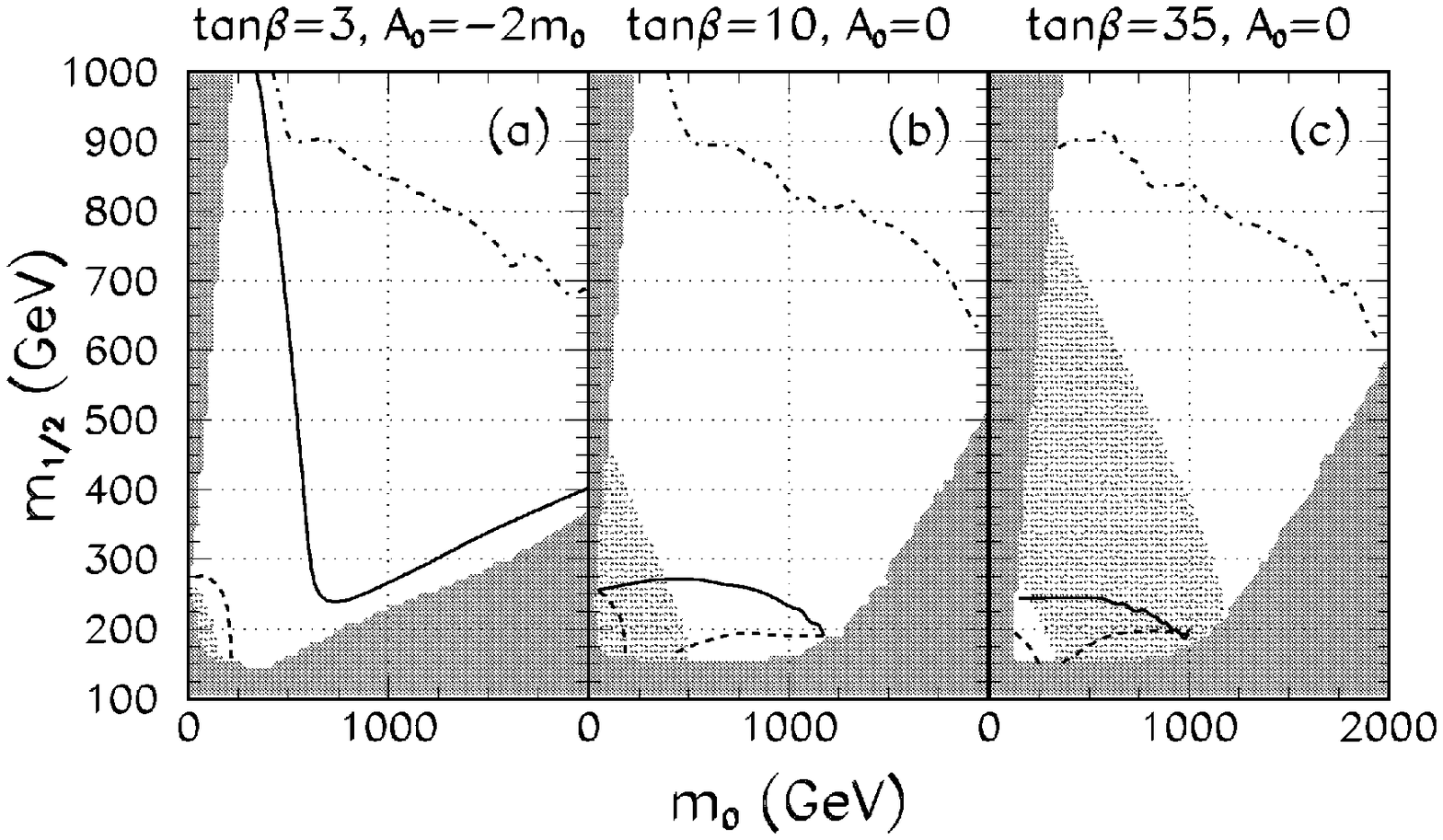}
\caption[]{
A plot of $m_0\ vs.\ m_{1/2}$ parameter space in the mSUGRA model
for $\mu >0$ and {\it a}) $A_0=-2m_0$ and $\tan\beta =3$, {\it b}) $A_0=0$ and
$\tan\beta =10$ and {\it c}) $A_0=0$ and $\tan\beta =35$. The $2\sigma$
region favored by the E821 measurement is shaded with dots. The region
below the
solid contour has $m_h<113.5$ GeV. The region below the dashed
contour is accessible to Tevatron searches with 25 fb$^{-1}$ of integrated
luminosity, while the region below the dot-dashed contour is accessible via
LHC sparticle searches with 10 fb $^{-1}$ of integrated luminosity.}
\label{msugra}
\end{figure}
\begin{figure}
\dofig{5in}{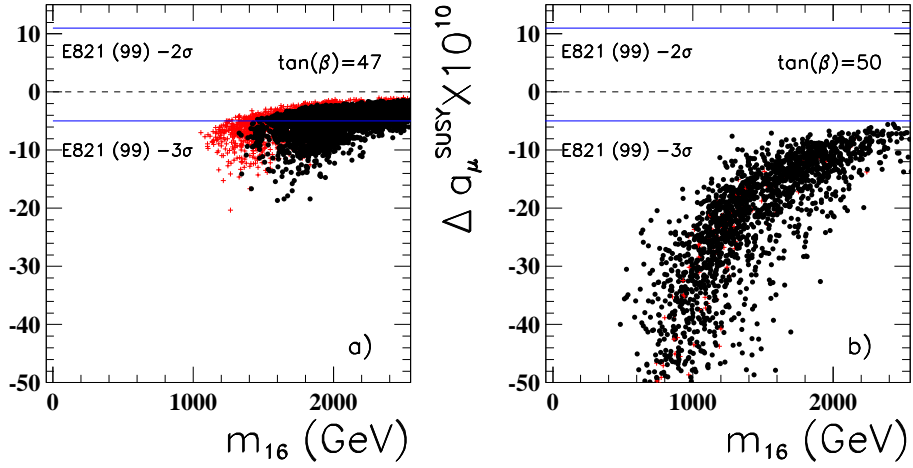}
\caption[]{
A plot of the $\Delta a_\mu^{SUSY}$ value for various Yukawa unified
$SO(10)$ models with {\it a}) $\tan\beta =47$, and $\mu <0$
and {\it b}) $\tan\beta =50$ and $\mu <0$.
We require $t-b-\tau$ Yukawa unification at $M_{GUT}$ at the 5\% level.
The pluses are valid solutions but have some sparticle or Higgs masses
in conflict with LEP2 constraints. }
\label{so10}
\end{figure}
\begin{figure}
\dofig{5in}{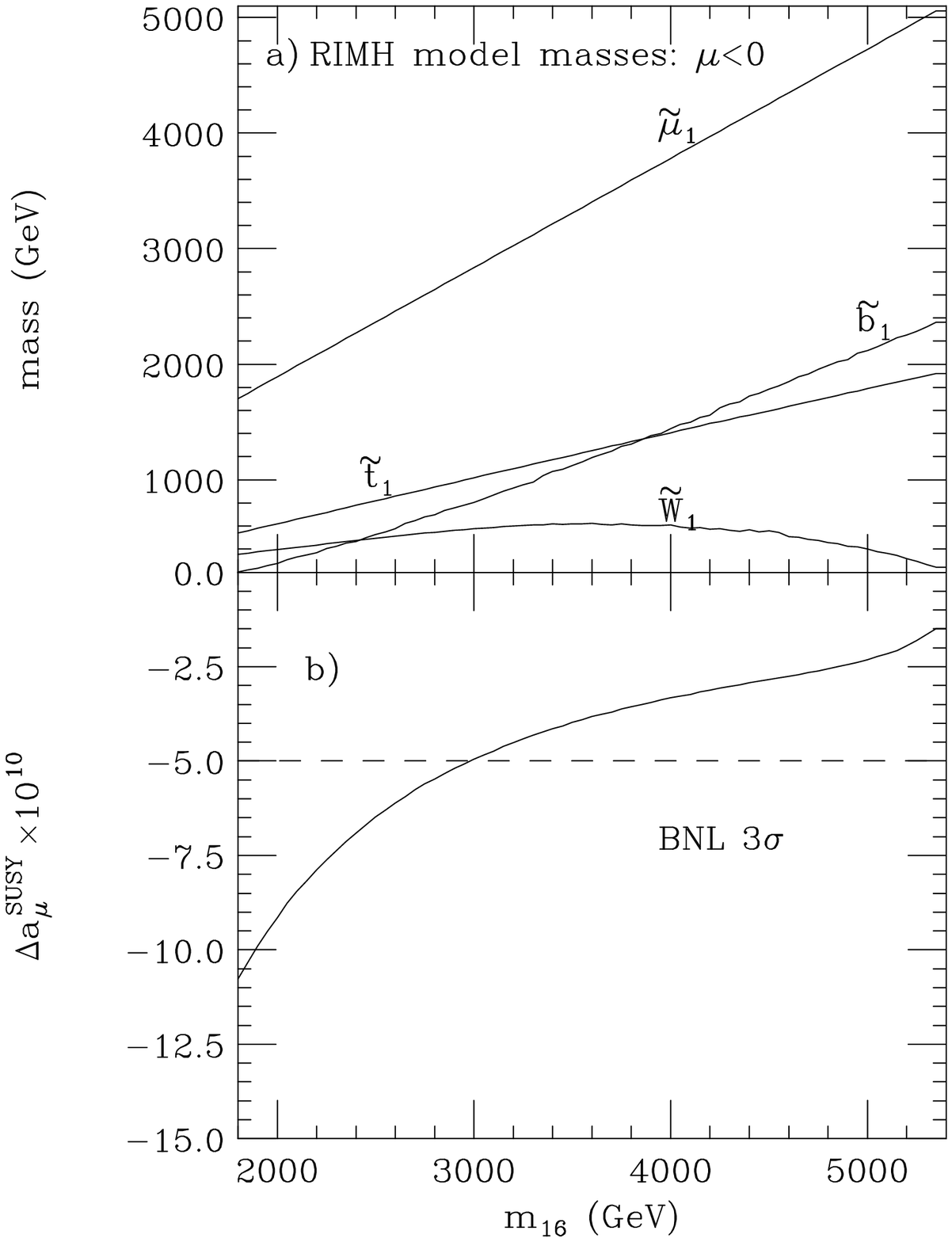}
\caption[]{
A plot of {\it a}) sparticle masses and
{\it b}) $\Delta a_\mu^{SUSY}$ versus $m_{16}$ for RIMH
models with $m_{1/2}=0.25 m_{16}$, $M_D=0.2 m_{16}$, $\tan\beta =50$,
$\mu <0$ and $M_N=1\times 10^7$ GeV.}
\label{imhn}
\end{figure}
\begin{figure}
\dofig{6.5in}{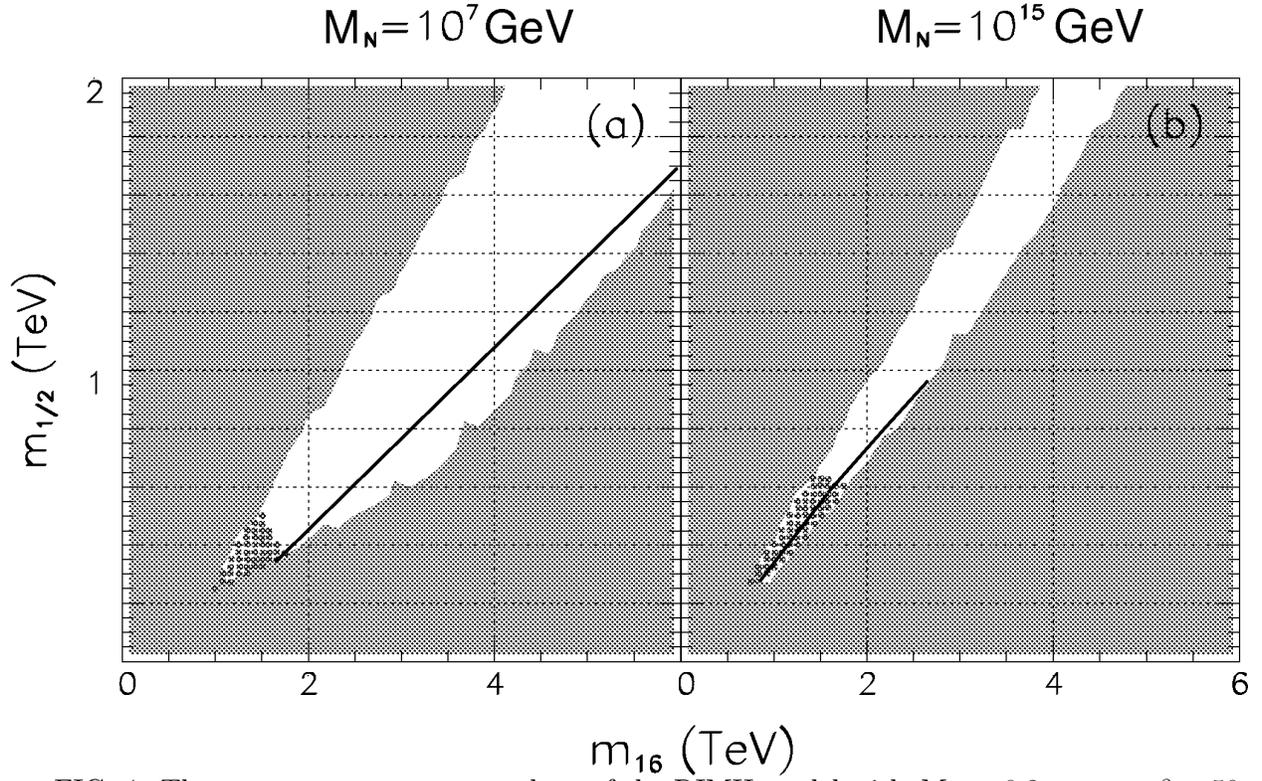}
\caption[]{The $m_{16}-m_{1/2}$ parameter plane of the RIMH model
with $M_D=0.2 m_{16}$, $\tan\beta =50$,
$\mu >0$, and {\it a})~$M_N=1\times 10^7$ GeV and {\it b})~$M_N =
10^{15}$~GeV.The shaded region is excluded by the theoretical constraints
discussed in the text. To the right of the solid line, the crunch
factor $S > 4$. In the region shaded with dots, the SUSY contribution to
$a_{\mu}$ is within $2\sigma$ of the central value obtained by the E821 experiment.}
\label{imhpos}
\end{figure}
\begin{figure}
\dofig{5in}{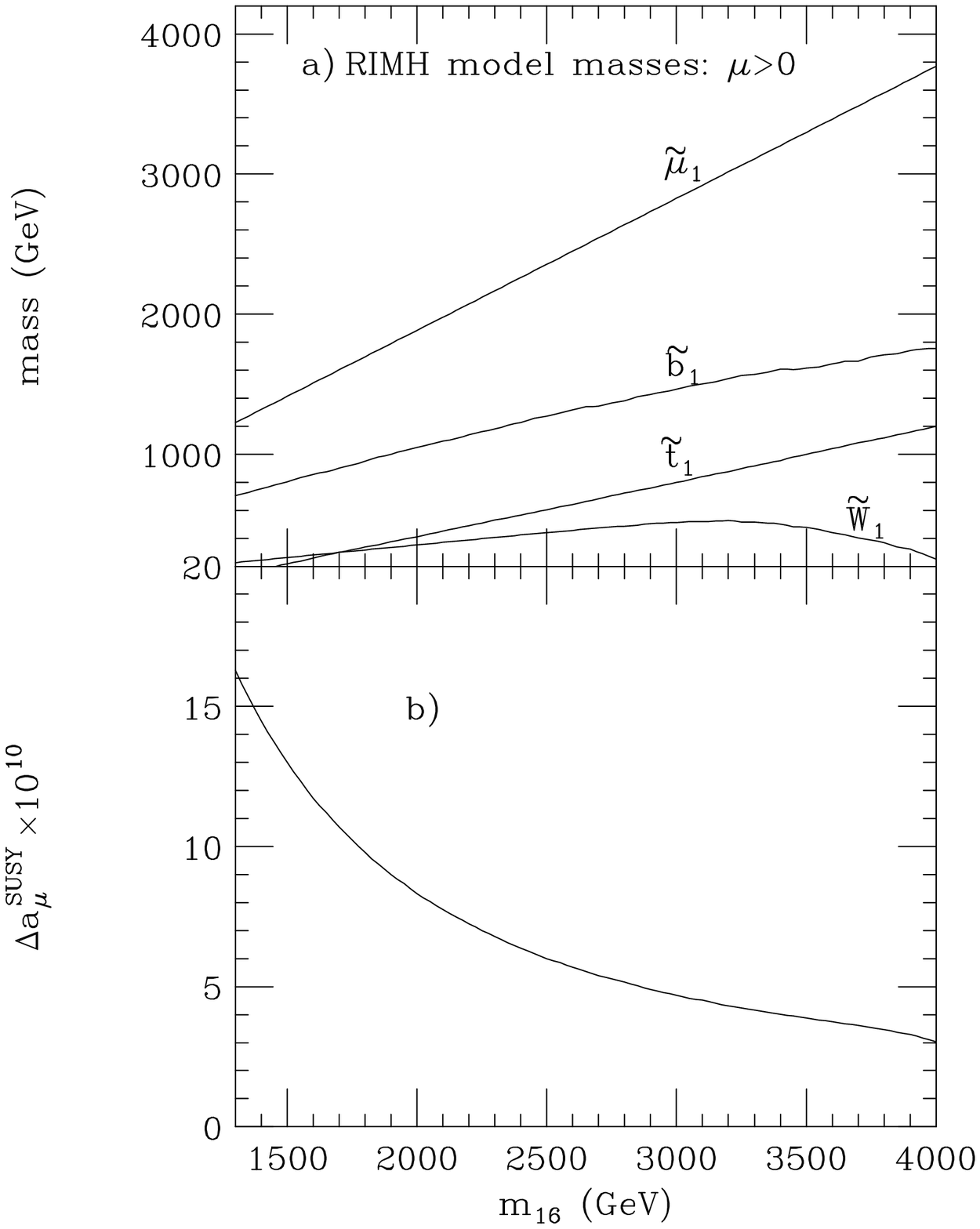}
\caption[]{
A plot of {\it a}) sparticle masses and
{\it b}) $\Delta a_\mu^{SUSY}$ versus $m_{16}$ for RIMH
models with $m_{1/2}=0.22 m_{16}$, $M_D=0.2 m_{16}$, $\tan\beta =50$,
$\mu >0$ and $M_N=1\times 10^7$ GeV.}
\label{imhp}
\end{figure}
\begin{figure}
\dofig{6.5in}{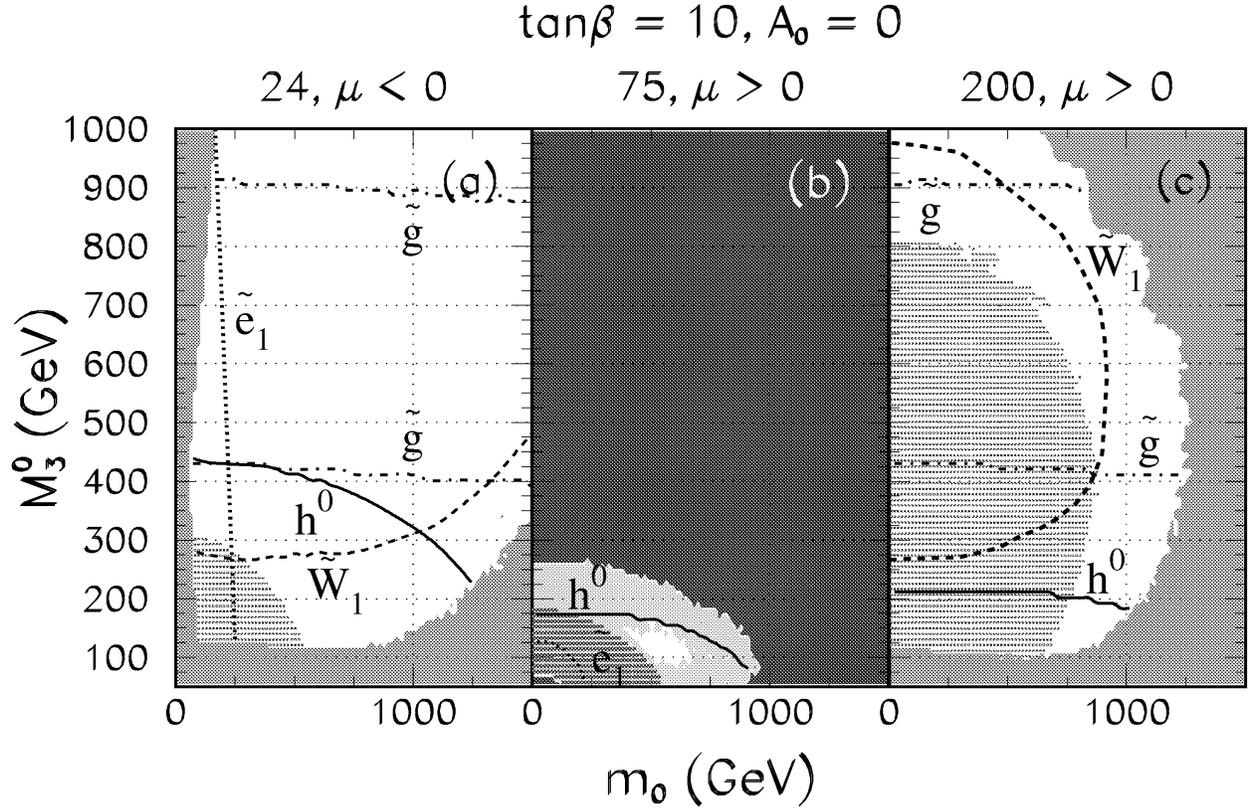}
\caption[]{A plot of the $m_0\ vs.\ M_3^0$ parameter plane of the $SU(5)$
model with non-universal gaugino masses discussed in Sec.~IID for the
case where the superfield $\Phi$ transforms as a {\it a})~{\bf 24}, {\it
b})~{\bf 75}, and {\it c})~{\bf 200} dimensional representation of
$SU(5)$. We have
fixed $A_0=0$ and $\tan\beta=10$. The shaded region in frames {\it
a}) and {\it c}) is excluded by
experimental and theoretical constraints discussed in the text. In frame
{\it b}), however, the dark shaded region corresponds to just the
theoretical constraints, while in the light shaded region $m_{\tw_1} <
85$~GeV. In the white region around $m_0 \sim 500$~GeV and
$M_3^0=125$~GeV, the chargino is between 85 and 100~GeV.The various lines labelled by a particle type are contours of
sparticle masses as discussed in the text. Finally, the $2\sigma$ region
favoured by the E821 experiment is shaded with dots.}
\label{gaugino1}
\end{figure}
\newpage
\begin{figure}
\dofig{6.5in}{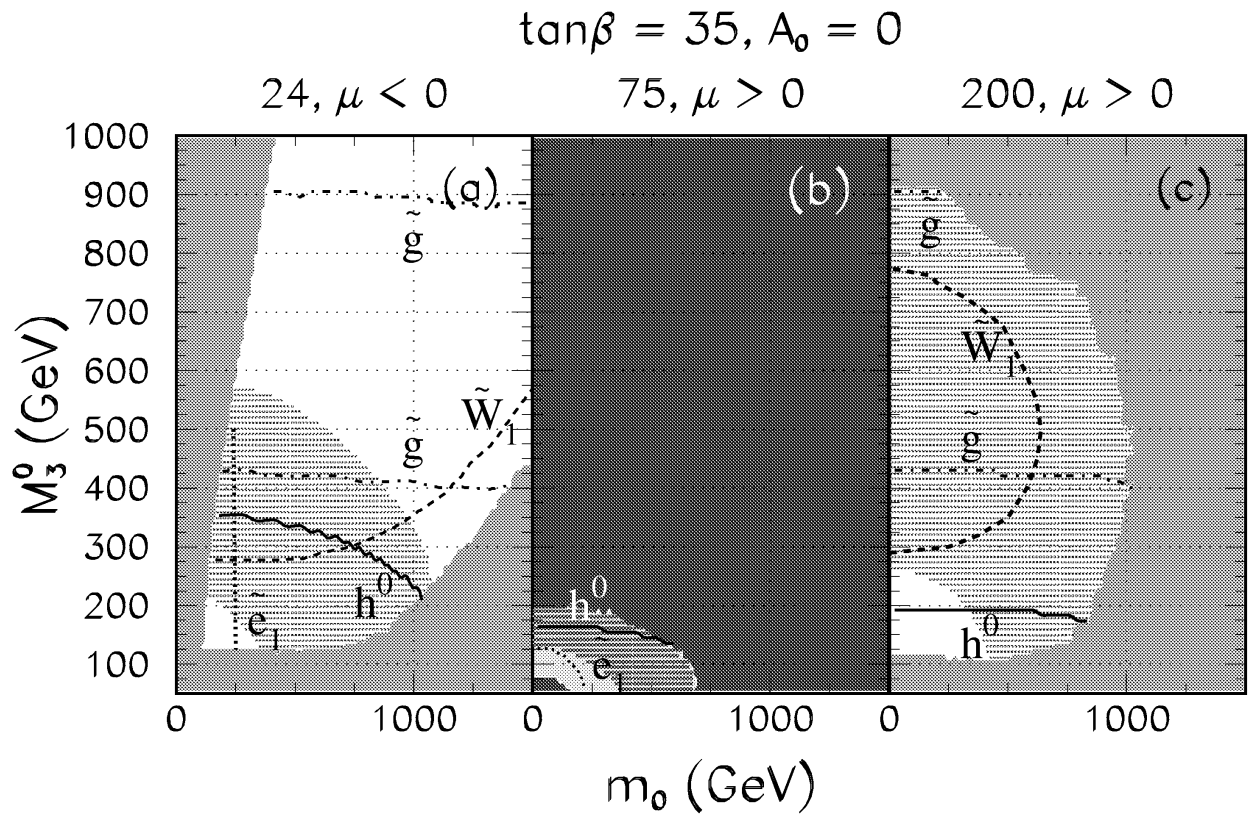}
\caption[]{The same as Fig.~\ref{gaugino1} except that $\tan\beta=35$.}
\label{gaugino2}
\end{figure}

\begin{figure}
\dofig{6.5in}{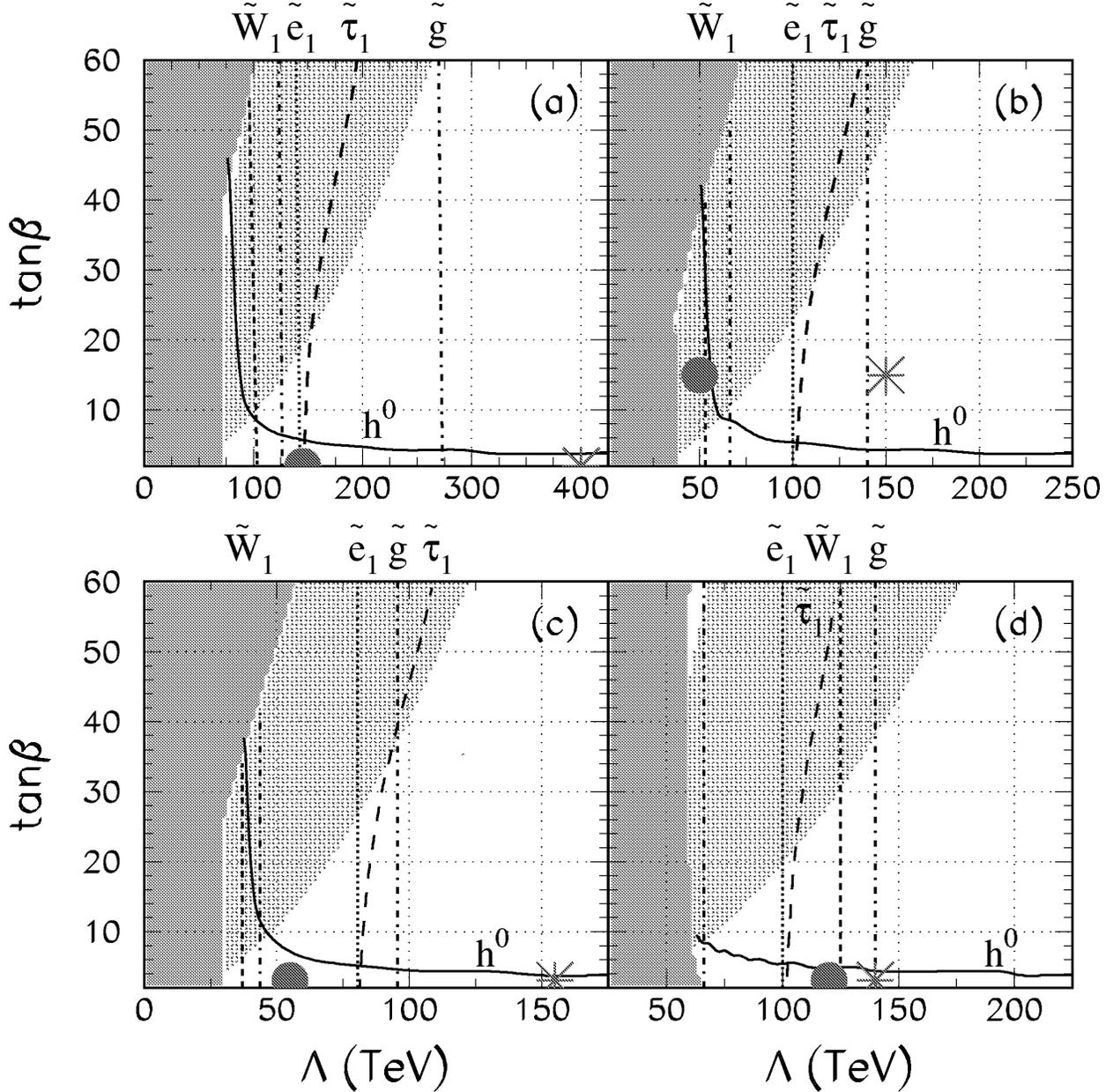}
\caption[]{
A plot of parameter space in the minimal GMSB model,
for $M=3\Lambda$, and $\mu >0$. In {\it a}), we take $n_5 =1$,
in {\it b}), $n_5=2$, in {\it c}) $n_5=3$ and in {\it d}),
$n_5=2$ but with $\mu =0.75 M_1$. The $2\sigma$ region favored by E821
is shaded with dots.
The reach in $\Lambda$ of the CERN LHC for particular model lines
with specific $\tan\beta$ values
is indicated by asterisks. For instance, in frame {\it a}) the model
line studied had $tan\beta=2$ and the reach extending to $\Lambda$
beyond 400~TeV. The large dots denote the corresponding reach of a
luminosity upgrade of the Tevatron that yields 25~$fb^{-1}$ of
integrated luminosity.}
\label{gmsb}
\end{figure}
\begin{figure}
\dofig{6.5in}{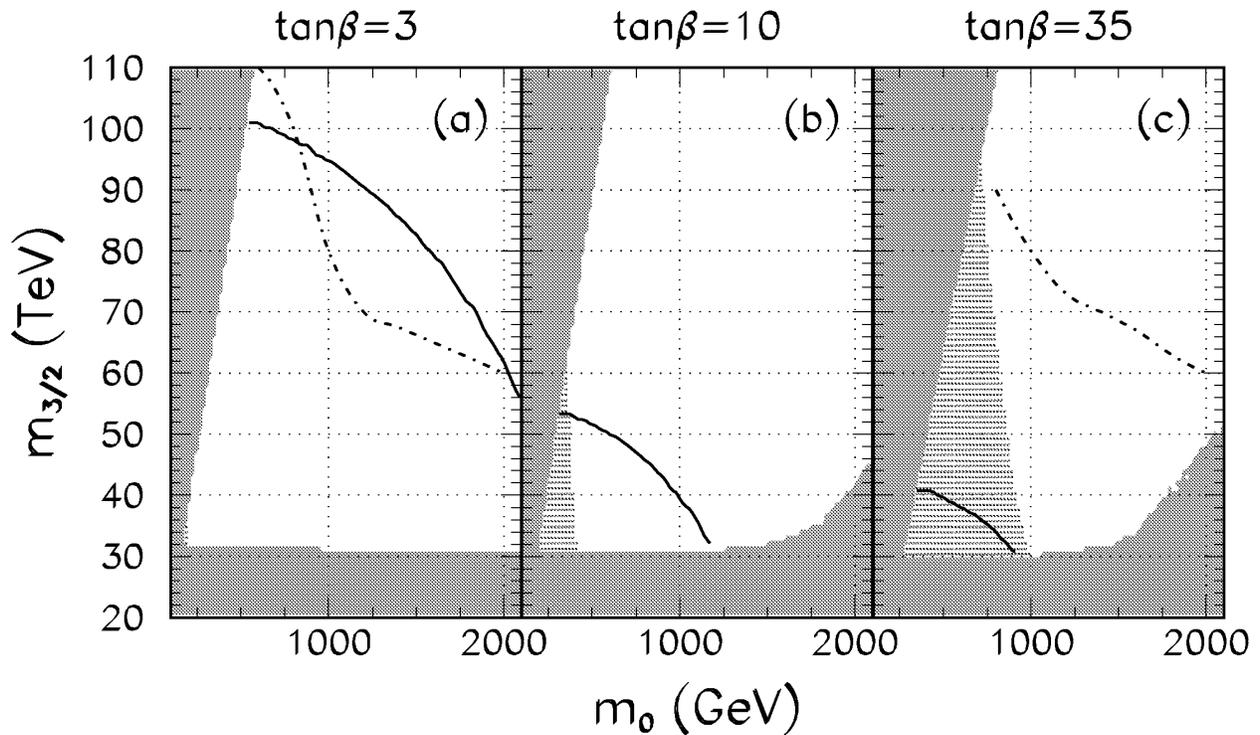}
\caption[]{
A plot of parameter space in the minimal AMSB model,
for $\mu <0$ and {\it a}) $\tan\beta =3$, {\it b}) $\tan\beta =10$  and
{\it c}) $\tan\beta =35$. In {\it a}), the solid contour denotes
where $m_h=105$ GeV, while in {\it b}) and {\it c}) it denotes where
$m_h=113.5$ GeV. The $2\sigma$ region favored by E821 is shaded with dots.
The dot-dashed contours denote the boundary of the region that will be
probed by the CERN LHC with 10 fb$^{-1}$
of integrated luminosity.}
\label{amsb}
\end{figure}
\begin{figure}
\dofig{5in}{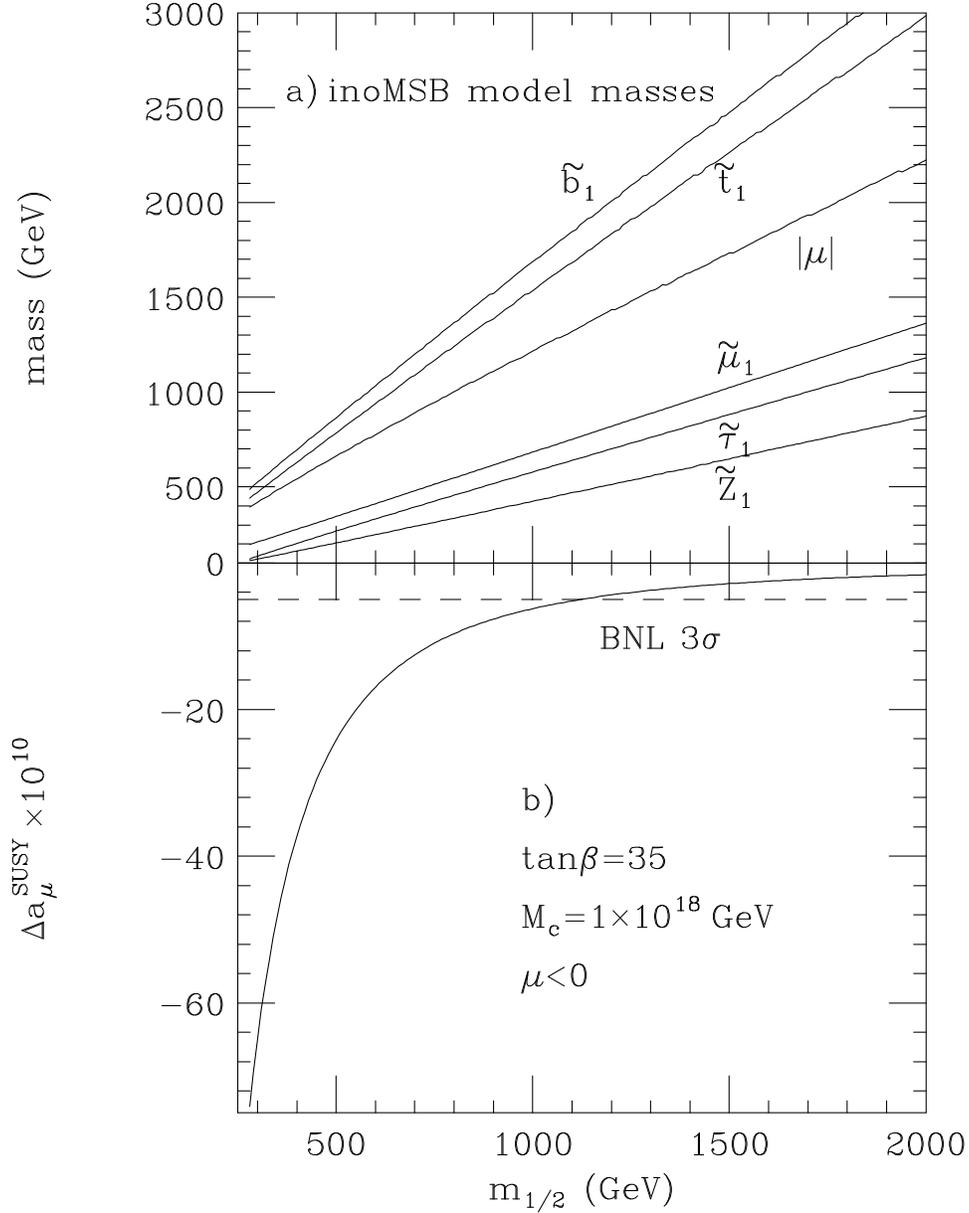}
\caption[]{
A plot of {\it a}) sparticle masses and {\it b})
$\Delta a_\mu^{SUSY}$ versus $m_{1/2}$ in the minimal gaugino-mediated
SUSY breaking model, for $\tan\beta =35$, and $\mu <0$. We assume
$SU(5)$ unification between $M_c$ and $M_{GUT}$.}
\label{inomed}
\end{figure}

\end{document}